\documentstyle[12pt]{article}
\begin{document}
 
\begin{titlepage}
\begin{center}
{\bf PHYSICAL MASSES AND THE VACUUM EXPECTATION VALUE}\\
{\bf OF THE HIGGS FIELD}\\
\end{center}

\begin{center}
Hung Cheng\\
Department of Mathematics, Massachusetts Institute of Technology\\
Cambridge, MA  02139, U.S.A.\\
\end{center}
\begin{center}
and
\end{center}
\begin{center}
S.P. Li\\
Institute of Physics, Academia Sinica\\
Nankang, Taipei, Taiwan, Republic of China\\
\end{center}
%\end{titlepage}

\vskip 3 cm
\begin{center}
{\bf Abstract}
\end{center}
By using the Ward-Takahashi identities in the Landau gauge, we derive exact  
relations between particle masses and the vacuum expectation value of
the Higgs field in the Abelian gauge field theory with a Higgs meson.

%\end{abstract}
\end{titlepage}

\newpage
Despite the pioneering work of 't Hooft and Veltman${^1}$
 on the renormalizability of gauge field theories with a spontaneously broken vacuum symmetry, a number of questions remain open${^2}$. 
In particular, the number of renormalized parameters in these theories exceeds
that of bare parameters.  For such theories to be renormalizable, there must
exist relationships among the renormalized parameters involving no infinities.

As an example, the bare mass $M_0$ of the gauge field in the Abelian-Higgs
theory is related to the bare vacuum expectation
value $v_0$ of the Higgs field by
$$
M_0 = g_0 v_0 ,
\eqno(1)
$$
where $g_0$ is the bare gauge coupling constant in the theory.  We shall show
that
$$
M =  \sqrt { \frac{D_T(0)}{D_T(M^2)} } {\hskip 0.2 cm} g(0) v .
\eqno(2)
$$
In (2), $g(0)$ is equal to $g(k^2)$ at $k=0$, with the running renormalized gauge coupling constant defined
in (27), and $v$ is the renormalized
vacuum expectation value of the Higgs field defined in (29) below.

Consider the Abelian-Higgs model with the Langrangian density given by
$$
L = - \frac{1}{4} F_{\mu\nu} F^{\mu\nu} + (D_{\mu}\phi)^+ (D^\mu \phi) +
 \mu^{2}_{0} \phi^{+}\phi - \lambda_0(\phi^{+}\phi)^2,
\eqno {(3)}
$$
with
$$
D_\mu \phi \equiv (\partial_\mu + ig_0 V_\mu)\phi, 
$$
and
$$
F_{\mu\nu} \equiv \partial_\mu V_\nu - \partial_\nu V_\mu.
$$
In the above, $V_{\mu}$ and $\phi$ are the gauge field and the Higgs field
respectively.  The subscript (0) of the constants in
(3) signifies that these constants are bare constants.  As usual, we shall put
$$
\phi = \frac{v_0 + H + i \phi_2}{\sqrt{2}}
\eqno{(4)}
$$
where
$
v_0 \equiv {\mu_0}/{\sqrt{\lambda_0}}.
$

To canonically quantize the theory given by this Lagrangian, we add 
a gauge-fixing term and the associated ghost terms to the Lagrangian.  The
effective Lagrangian obtained is
$$
L_{eff} \equiv L - \frac{1}{2\alpha}\ell^2 -
i(\partial_{\mu}\eta)(\partial^{\mu}\xi)+i \alpha M^2_0 \eta\xi + i \alpha g_0
M_0 \eta\xi H, 
\eqno(5)
$$
In (5)
$$
\ell \equiv \partial_\mu V^\mu-\alpha M_0 \phi_2,
\eqno{(6)}
$$
$\alpha$ is a constant
and $\xi$ and $\eta$ are ghost fields.  The Lagrangian in (5) is the effective
Lagrangian in the $\alpha$-gauge.
It is invariant under the following BRST variations:
$$
\delta V_\mu = \partial_\mu\xi, \hspace{.25in}\delta H= g_0 \xi\phi_2,
\hspace{.25in} \delta\phi_2 = - g_0 \xi(v_0+H),
$$
$$
\delta i\eta = \frac{1}{\alpha}\ell, \hspace{.25in} \delta\xi=0 .
\eqno(7)
$$

There exists a specific formula relating the vacuum state $|0>$ in this
effective theory$^3$ to that in the original gauge theory.
This vacuum state satisfies  
$$
Q|0> = 0.
\eqno{(8)}
$$
In (8), $Q$ is the BRST charge, the commutator (anticommutator) of which
with a bosonic (Grassmann) field is the BRS variation for this field, e.g., 
$$
[iQ, V_\mu]= \delta V_\mu .
\eqno{(9)}
$$
The Ward-Takahashi identities are easily derived from (8) and (9).  For example,
we have, as a consequence of (8)$^4$,
$$
<0|[iQ, Ti\eta(x)\phi_2(y)]_+|0 > = 0,
\eqno{(10)}
$$
where $T$ signifies time-ordering.  However, the anticommutator above is 
indeed the BRST variation of $Ti\eta(x)\phi_2(y)$.  Thus
we obtain, by (9), the Ward-Takahashi identity
$$
<0|T(\frac{1}{\alpha}\partial_\mu V^{\mu}(x)-
M_0\phi_{2}(x))\phi_2(y)+Ti\eta(x)\xi(y)(M_0+g_0 H(y))|0>=0 .
\eqno{(11)}
$$
All other Ward-Takahashi identities can be derived in a similar way.
It is particularly convenient to study these Ward-Takahashi identities in the limit $\alpha\rightarrow 0$, i.e.,
in the Landau gauge. 

 The propagators for $H, \phi_2$, and the ghost field will be
denoted by 
$$
\frac{i
Z_H(k^{2},\alpha)}{k^2-2\mu^2}\, ,\,
 \frac{i Z_{\phi_{2}}(k^2,
\alpha)}{k^2}\, , \,
{\rm and}\,\,
\frac{iZ_\eta(k^2,\alpha)}{k^2}
$$
respectively, where $k$ is the momentum of the particle and
${\sqrt 2} \mu$ is the physical mass of the Higgs meson.
The propagator for the vector meson will be denoted as:
$$
-i(g^{\mu\nu}- \frac{k^\mu k^\nu}{k^2}) \frac{Z_T(k^2,\alpha)}{k^2-M^2} - i\alpha
\frac{k^\mu k^\nu}{k^2} \frac{Z_L(k^2,\alpha)}{k^2},
\eqno{(12)}
$$
where the subscripts $T$ and $L$ signify transverse and longitudinal
respectively, and where $k^2$ is really $k^2+i\epsilon$, with $\epsilon$
positive and infinitesimal. In the lowest order of perturbation, all the Z functions 
are equal to unity.  Finally, there is the propagator
$$
\int dx e^{ik.x}<0|TV^\mu(x)\phi_2(0)|0> \equiv \alpha \frac{k^\mu M_0}{k^2}
\frac{Z_{L2}(k^2\alpha)}{k^2},
\eqno{(13)}
$$
which is not zero since $\phi_2$ mixes with the longitudinal component of
$V^\mu$ as soon as the interaction is turned on.  In the unperturbed order, 
$Z_{L2}=0$ as this mixing is due to the interaction of $\phi$ with other
fields.  

In the Landau gauge, the longitudinal part of (12) vanishes.  The propagator in (13) also vanishes at
$\alpha=0$.  In addition, the ghost field is decoupled from the other fields at
$\alpha=0$.  Thus $Z_\eta(k^2,0)=1$.  We shall take advantage of these simplifications occuring in the
Landau gauge. 

Let us take the Fourier transform of (11) and then take the limit
$\alpha\rightarrow 0$.  We get, after some algebra 
$$
Z_{L2}(k^2) + Z_{\phi_{2}}(k^{2}) = Z, 
\eqno{(14)}
$$
where
$$
Z \equiv \frac{1}{v_0} <0|(v_0+H)|0>.
\eqno{(15)}
$$
In the above, $Z_{L2}(k^2)$ is $Z_{L2}(k^2,0)$, and similarly for
$Z_{\phi_2}(k^2)$.  Note that $Z$ is the ratio of the quantum expectation
value of $\phi$ with the classical value of the vacuum state of $\phi$
and is independent of $k^2$.

Next we turn to the Ward-Takahashi identity obtained by setting
$$
<0|[iQ, Ti\eta(\frac{1}{\alpha}\partial_\nu V^\nu-M_0\phi_2)]_+|0>
\eqno(16)
$$
to zero.  
Taking the limit $\alpha\rightarrow 0$ for (16)
requires a little care, as there are terms in (16) which are of the order of
$\alpha^{-1}$.  These terms will cancel and we are left with the terms
which are finite as we take the limit $\alpha\rightarrow 0$.  
We get
$$
\lim_{\alpha\rightarrow 0} \frac{1-Z_{L}(k^2,\alpha)}{\alpha} +
\frac{2M^2_0}{k^2}Z_{L2}(k^2) + \frac{M^2_0}{k^2}Z_{\phi{_2}}(k^2)=0.
\eqno{(17)}
$$

Let us denote the 1PI (one-particle-irreducible) amplitude for the propagation
of $V^\nu$ to $V^\mu$ by
$$
g^{\mu\nu}A(k^2,\alpha)+k^\mu k^\nu B(k^2,\alpha)
$$
which can be written as
$$
(g^{\mu\nu}-\frac{k^\mu
k^\nu}{k^2})A(k^2,\alpha)+\frac{k^\mu k^\nu}{k^2}[A(k^2,\alpha)+k^2
B(k^2,\alpha)]. 
\eqno{(18)}
$$
We also denote the 1PI amplitude for the propagation of $\phi_2$ to $\phi_2$ by
$
C(k^2,\alpha),
$
and the 1PI amplitude for the propagation of $V^\mu$ into $\phi_2$ by
$
\displaystyle i\frac{k^\mu}{M_0}D(k^2,\alpha).
$
By writing out the unrenormalized propagators in terms of $A,B,C$ and 
$D$, we find
that the mixing of $\phi_2$ with the longitudinal component of $V^\mu$ is
given by 
$$\displaystyle
Z_L(k^2,\alpha) = 1+ \alpha[\frac{M^2_0+A+k^2B}{k^2} -
\frac{D^2}{M^2_0(k^2+C)}] + 0(\alpha^2),
\eqno(19a)
$$
$$
 \displaystyle
Z_{\phi{_2}}(k^2,\alpha) = \frac{k^2}{k^2+C} + 0(\alpha),
\eqno(19b)
$$
and
$$
Z_{2L}(k^2,\alpha) = - \frac{D}{k^2+C} \frac{k^2}{M^2_0} + 0(\alpha).
\eqno{(19c)}
$$
From the last two equations of (19), we get, setting $\alpha\rightarrow 0$,
$$
\frac{M_0^2}{k^2} \frac{Z^2_{L2}(k^2)} {Z_{\phi_{2}}(k^2)} =
\frac{D^2(k^2)}{k^2+C(k^2)} \frac{1}{M_0^2}.
\eqno{(20)}
$$
Thus, the equation (19a) gives
$$
\lim_{\alpha\rightarrow 0} \frac{Z_L(k^2,\alpha)-1}{\alpha} =
\frac{M^2_0+A(k^2)+k^2B(k^2)} {k^2} - \frac{M^2_0}{k^2}
\frac{Z^2_{L_{2}}(k^2)}{Z_{\phi_{2}}(k^2)}.
\eqno{(21)}
$$
Substituting the above into the Ward-Takahashi identity (17), we get
$$
M^2_0 + A(k^2)+ k^2B(k^2) = M^2_0 \frac{Z^2}{Z_{\phi_{2}}(k^2)},
\eqno{(22)}
$$
where (14) has been utilized.

The unrenormalized propagator for the transverse components of $V^{\mu}$ at
$\alpha=0$ is
$$\displaystyle
\frac{-i(g^{\mu\nu} -k^\mu k^\nu/k^2)}{k^2-M^2_0-A(k^2)} .
$$
Thus
$$
Z_T(k^2) = \frac{k^2-M^2}{k^2-M^2_0-A(k^2)},
\eqno{(23)}
$$
or
$$
M^2_0 +A(k^2) = k^2 + \frac{M^2-k^2}{Z_T(k^2)}.
\eqno(24)
$$
>From (22) and (24), we get
$$
k^2 + \frac{M^2-k^2}{Z_T(k^2)} + k^2B(k^2) = M^2_0 \frac{Z^2}{Z_{\phi_{2}}(k^2)}.\eqno{(25)}
$$
Finally, we set $k^2=0$, obtaining
$$
M^2 = M^2_0 \frac{Z^2 Z_T(0)} {Z_{\phi_{2}}(0)},
= \frac{Z^2 v^2_0} {Z_{\phi_{2}}(0)}g^2_0 Z_T(0).
\eqno{(26)}$$
We define the running renormalized coupling constant $g(k^2)$ as
$$
g(k^2)\equiv g_0\sqrt{Z_T(k^2)}
\eqno{(27)}
$$

Also, we define the vacuum expectation value of the renormalized scalar field
as
$$
v \equiv <0| \frac{v_0+H}{\sqrt{Z_{\phi_{2}}(0)}}|0>.
\eqno{(28)}
$$
By (15), we have
$$
v = \frac{v_0 Z}{\sqrt{Z_{\phi_{2}}(0)}}.
\eqno(29)
$$
From (22), (23) and (24), we get (2).

Finally, we mention that it is also possible to relate the mass of a fermion
to the vacuum expectation value of the Higgs field.  Similar to (1), we have,
in the classical theory,
$$
m_0=f_0v_0,
$$
where $m_0$ is the bare mass of the fermion, and $f_0$ is the bare Yukawa
coupling constant.  In the quantum theory, we have$^5$, similar to (2)
$$
m=fv/[1+a_r(m^2)].
\eqno{(30)}
$$
where $a_r(p^2)$ is an invariant amplitude in the renormalized fermion
propagator. 

\newpage

\begin{center}
{\bf References}
\end{center}
\vskip .25in

\begin{description}

\item{1.  G. t'Hooft and M. Veltman, \underline{Nuclear Physics},{\bf B44}, 189, (1972).}

\item{2.  See, for example, the review article by G. Bonneau,
\underline{Int'l Journal of Modern Physics A}, {\bf Vol. 5}, No. 20, 3831 (1990).}

\item{3.  H. Cheng and E.C. Tsai, \underline{Physics Review}, {\bf D40}, 1246,
(1989).  See also H. Cheng in \underline{Physical and Nonstandard Gauges},
edited by Gaigg et. al, Springer-Verlag (1989).}

\item{4. The method we use to obtain the Ward-Takahashi identities here was shown to one of us(H. Cheng) by E.C. Tsai in a private communication.}
\item{5.  H. Cheng and S.P. Li, ``How to Renormalize a Quantum Gauge Field
Theory with Chiral Fermions'', (submitted for publication, 1996).}
\end{description}

\end{document}